\documentclass[pdflatex,sn-mathphys-num]{sn-jnl}

\usepackage{graphicx}%
\usepackage{multirow}%
\usepackage{amsmath,amssymb,amsfonts}%
\usepackage{amsthm}%
\usepackage{mathrsfs}%
\usepackage[title]{appendix}%
\usepackage{xcolor}%
\usepackage{textcomp}%
\usepackage{manyfoot}%
\usepackage{booktabs}%
\usepackage{algorithm}%
\usepackage{algorithmicx}%
\usepackage{algpseudocode}%
\usepackage{listings}%
\usepackage{cleveref}
\usepackage{caption}
\usepackage{subcaption}
\usepackage{mathtools}
\usepackage{xspace}
\usepackage{tabularx}

\theoremstyle{thmstyleone}%
\theoremstyle{thmstyletwo}%

\theoremstyle{thmstylethree}%

\newcommand{\bnfdef}{\ensuremath{\Coloneqq}}
\newcommand{\bnfalt}{\ensuremath{\mid}\xspace}

\newcommand{\ronefull}{R1: Insertions adjacent to ranges are ambiguous}
\newcommand{\rtwofull}{R2: Sorting permutes referenced data}
\newcommand{\rthreefull}{R3: Moving cells changes references}
\newcommand{\rfourfull}{R4: Names do not always move alongside data}
\newcommand{\rfivefull}{R5: References are relative by default}

\begin{document}

\title[Article Title]{Kale: A Transformation-Safe Spreadsheet System}


\author*[1]{\fnm{Michael}\sur{Coblenz}}\email{mcoblenz@ucsd.edu}

\author[1]{\fnm{Jacob} \sur{Yim}}\email{j1yim@ucsd.edu}

\author[2]{\fnm{Ajinkya}\sur{Bokade}}\email{acbokade@gmail.com}
\author[3]{\fnm{Mounika}\sur{Padala}}\email{mounikapadala11@gmail.com}
\author[4]{\fnm{Julia}\sur{Epshtein}}\email{jepshtein@umass.edu}
\author[5]{\fnm{Priyanka}\sur{Bhatia}}\email{b.priyanka0210@gmail.com}
\author[6]{\fnm{Piyush}\sur{Chauhan}}\email{piyushchauhan1004@gmail.com}
\author[3]{\fnm{Simran}\sur{Gill}}\email{gillsimu98@gmail.com}
\author[1]{\fnm{Aniket}\sur{Gupta}}\email{ang042@ucsd.edu}
\author[1]{\fnm{Grishma}\sur{Gurbani}}\email{grishmagurbani1998@gmail.com}
\author[3]{\fnm{Vaibhav}\sur{Khetan}}\email{vaibhavkhetan2703@gmail.com}
\author[1]{\fnm{Arushi}\sur{Munjal}}\email{amunjal@ucsd.edu}
\author[1]{\fnm{Jeffery}\sur{Tung}}\email{j7tung@ucsd.edu}
\author[1]{\fnm{Joanna}\sur{Yang}}\email{joy002@ucsd.edu}

\affil[1]{\orgname{University of California, San Diego}}
\affil[2]{\orgname{Salesforce, Inc}}
\affil[3]{\orgname{Apple, Inc}}
\affil[4]{\orgname{University of Massachusetts, Amherst}}
\affil[5]{\orgname{Meta, Inc}}
\affil[6]{\orgname{Eukarya, Inc}}


\abstract{Spreadsheet formulas can refer to rectangular ranges of arbitrary size. When a user changes the structure of a referenced table, the spreadsheet system updates the references to refer to a new range. Unfortunately, this new range may differ from the user's expectations, introducing bugs in spreadsheets. We describe a user study showing that standard reference semantics are error-prone, resulting in significant risk to users. We introduce Kale, a prototype system that eliminates the risk of inserting these kinds of bugs by restricting the kinds of references that can be expressed. We show that Kale can be used effectively by users to complete tasks that are error-prone in traditional spreadsheet systems. Finally, we describe a corpus study that evaluates the extent to which the reference restrictions in Kale might have implications on users.
}




\maketitle
\section{Introduction}
\label{sec:intro}

Spreadsheets may well represent the world's most popular programming language~\cite{Scaffidi2005:Estimating}, with likely hundreds of millions of users worldwide~\cite{Gislason2018:Excel}. They are used for essential work in nearly every industry, including business, finance, economics, and science. Unfortunately, many spreadsheets are buggy, potentially leading users to incorrect conclusions. Various studies have shown that about 95\% of spreadsheets contain errors~\cite{Panko2015:What}. Spreadsheet errors have resulted in significant economic and scientific costs. For example, J.P Morgan Chase's 2012 ``London Whale'' debacle stemmed from a spreadsheet error, costing the company \$6.2B~\cite{Chase2013:Report}. Approximately one fifth of genetics papers that included supplementary Excel gene lists were found to include erroneous gene name conversions~\cite{Ziemann2016:Gene}.

Most prior work on spreadsheet errors has focused on the possibility of entering incorrect formulas. In this work, we provide empirical evidence for a \emph{different} cause of spreadsheet errors that has not, as far as we are aware, been studied previously. When a spreadsheet table that is referenced by formulas is modified structurally (for example, by inserting a row), the referencing formulas are automatically updated by the spreadsheet system. For example, if a formula references cell B2 of a table, and the user inserts a row at the top of the table, the formula is then updated to refer to cell B3. This behavior is intended to preserve the semantics of the formula, since any data that was previously located in cell B2 will now be located in cell B3.

But do spreadsheet references actually refer to the \emph{data} in a cell, or do they refer to the \emph{geometry} of the cell? In the previous example, they refer to the data; if they referred to the geometry, they would still point to the same coordinates after the change. Consider the formula \texttt{SUM(B2:C3)}, which refers to a 2x2 rectangle of cells, and suppose that the user drags and drops cell C3 to a new location. The formula remains unchanged, but it may evaluate to a different value because cell C3 is now empty. In this case, the reference apparently referred to the \emph{geometry}, not the data. We refer to this problem as \emph{reference instability}: references are unstable under structural changes to spreadsheets, which can result in the introduction of bugs into previously-correct spreadsheets.

\begin{figure}
    \centering
    \includegraphics[width=\linewidth]{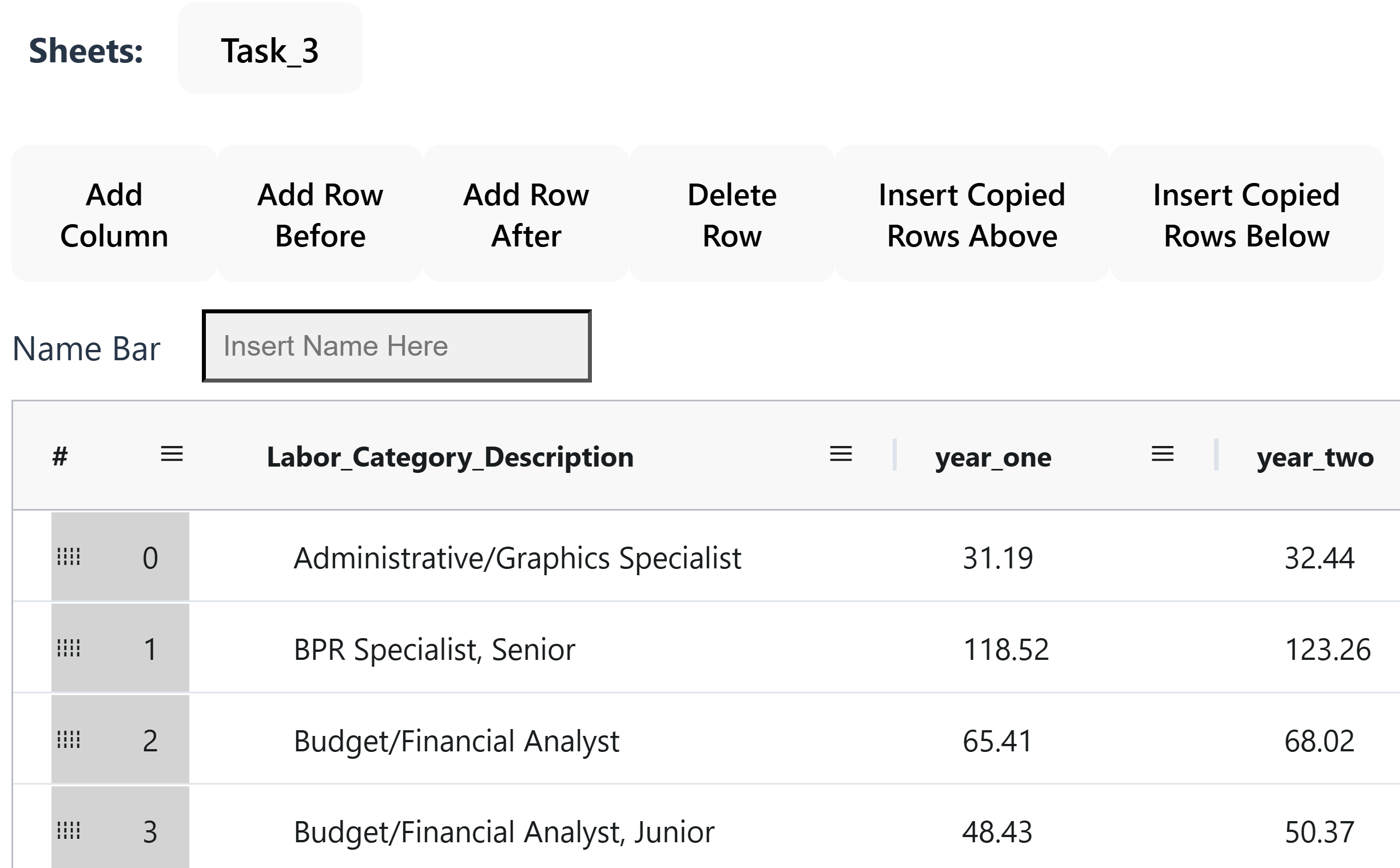}
    \caption{Kale, as seen in our Payroll task. A spreadsheet table is shown at the bottom; above are tabs showing available sheets and a set of buttons with available actions.}
    \label{fig:t3kale}
\end{figure}


Kale, shown in \Cref{fig:t3kale}, is a novel prototype spreadsheet system that improves safety by re-envisioning the syntax and semantics of spreadsheet references. In Kale, references can refer to individual cells, whole rows, and whole columns, but not spans of rows or columns, and not arbitrary rectangular ranges. In doing so, Kale avoids the reference instability that is fundamental to traditional spreadsheet systems.

Kale provides a key safety property not provided by traditional spreadsheets: \emph{preservation of referenced data through structural transformations}. That is, when rows or columns are inserted, removed, or permuted, formulas that reference data in affected cells will continue to refer to the same data after the transformation. Furthermore, when a formula itself is moved, it will refer to either the same data as before or the same relative offset according to the intent of the formula's author. This property does not hold in traditional spreadsheets.

Kale is designed to eliminate the risks posed by traditional spreadsheet systems: transformations can introduce bugs in formulas by breaking assumptions made by formula authors about which data formulas refer to. Each transformation presents a risk, which we index \textbf{R1}, \textbf{R2}, etc.

\textbf{\ronefull}. When rows or columns are inserted immediately above or below a referenced range, it can be unclear whether the reference should be updated to include the new rows or columns. If the new row or column represents another instance that is similar to the nearby ones, failing to include the new item represents an error. Sheets, Excel, and Numbers do not extend references to include new adjacent rows, but Excel emits a warning in the referencing cell.
 
\textbf{\rtwofull}. Sorting permutes rows in a table without updating any references to the sorted table. If an author intended a formula to refer to data in a row that the sort operation moves, then sorting breaks that formula.

\textbf{\rthreefull}. Moving rows or columns causes references that start or end at the moved row or column to be updated. For example, if a formula references \texttt{B2:B3} and the user moves row 3 below row 4, then the formula now refers to \texttt{B2:B5}, suddenly including the data in the row that was previously at index 4. In Excel, this operation is triggered by cut/paste; in Numbers and Sheets, by drag/drop.

\textbf{\rfourfull}. Cells can be given names. When structural changes occur, it can be unclear whether names will move. In Excel and Sheets, names move with values when cells are cut and pasted but not when cells are sorted.
 
\begin{figure}
    \centering
    
    \begin{subfigure}[t]{.45\linewidth}
        \centering
        \includegraphics[width=\linewidth]{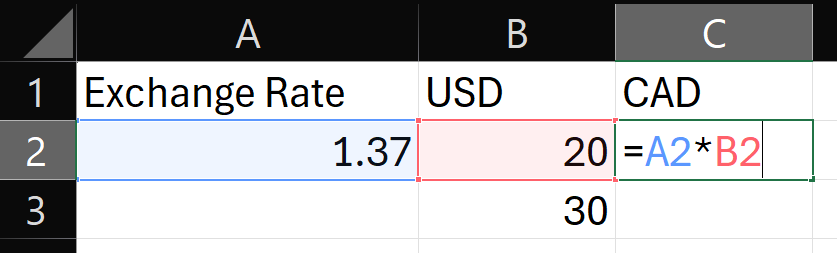}
        \caption{The user uses the exchange rate in cell A2 to convert the value in cell B2 to CAD.}
    \end{subfigure}
    \hfill
    \begin{subfigure}[t]{.45\linewidth}
        \centering
        \includegraphics[width=\linewidth]{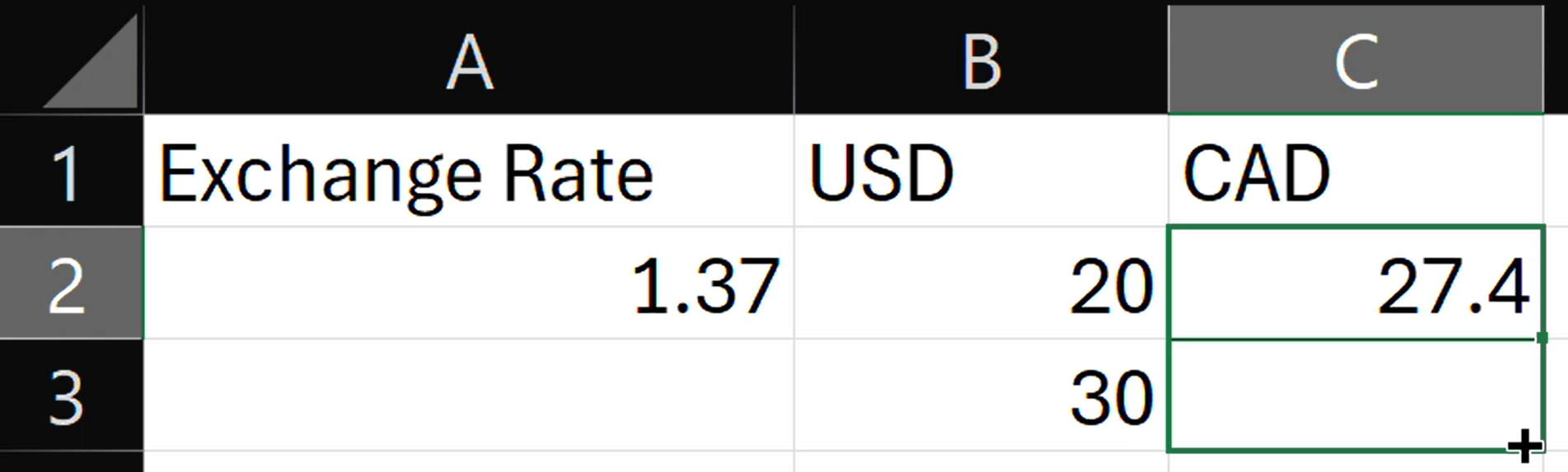}
        \caption{The user drag-fills the formula down.}
    \end{subfigure}
    
    \vspace{0.8em} 
    
    \begin{subfigure}[t]{.45\linewidth}
        \centering
        \includegraphics[width=\linewidth]{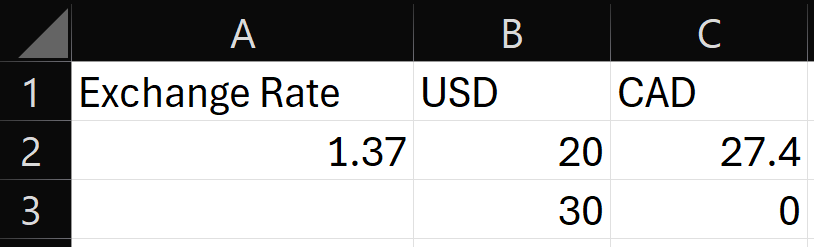}
        \caption{The value of the produced formula is unexpectedly 0.}
    \end{subfigure}
    \hfill
    \begin{subfigure}[t]{.45\linewidth}
        \centering
        \includegraphics[width=\linewidth]{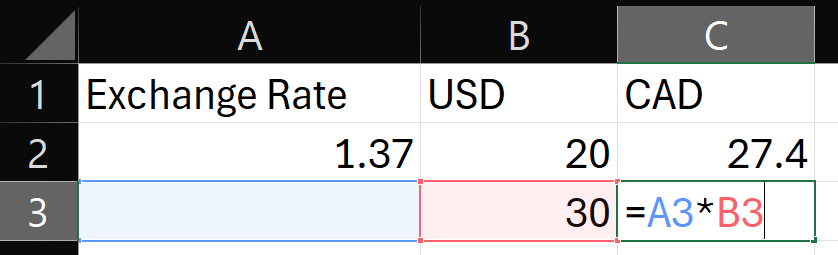}
        \caption{Upon closer inspection, the user sees that the new formula no longer references the intended exchange rate.}
    \end{subfigure}

    \caption{References are relative by default in Excel, leading to errors when users intend to use absolute references.}
    \label{fig:relreferror}
\end{figure}

\textbf{\rfivefull}. When a formula cell is copy-pasted or drag-filled, any resulting formulas are automatically adjusted to their new location. For instance, if a formula in cell B1 referencing cell A1 (the cell directly to the left) is drag-filled down to cell B2, the new formula now references cell A2. If the author intends this new formula to reference the \emph{same} cell as the copied formula, this behavior creates an error. A scenario demonstrating a variation of this error is shown in \Cref{fig:relreferror}. To avoid this error, the author would have to explicitly write absolute references; in Excel, Sheets, and Numbers, this is done by using the dollar sign symbol (\$) to mark absolute row and column references (e.g. \texttt{\$A\$1}).

Confusingly, although traditional spreadsheets distinguish between \emph{relative} and \emph{absolute} references, this distinction does not address risks R1-R4. In Excel, Sheets, and Numbers, the absolute vs. relative distinction \emph{only} affects which cells are referenced by a formula that is created by copy/paste or drag-fill; it has no influence on how a formula is updated when the structure of the referenced table is modified. For example, if a formula includes the absolute reference \texttt{\$B\$2} and the formula is copy/pasted one cell below its original location, the formula would still refer to \texttt{\$B\$2}. In contrast, if the reference had been relative (\texttt{B2}), it would then refer to \texttt{B3}. But both absolute and relative references are treated identically when row 2 is drag/dropped to a new location: the original formula might then refer to \texttt{\$B\$7}, for example, if the new location of the data is row 7. In Kale, the terms \emph{absolute} and \emph{relative} are re-defined so that they capture the semantics that a reference should have when a table is re-structured.

We describe a user study comparing Kale to traditional spreadsheets. We asked participants to complete four tasks using either Google Sheets or Kale. Each task is associated with a subset of the risks described in Section 1. The study focuses on assessing whether, when participants are given tasks that may break formulas, participants indeed transform their document in ways that introduce bugs. For participants who do introduce bugs, we assess whether they notice and are able to successfully fix them.

Because Kale restricts the kinds of references that can be expressed, there is a possibility that common problems that spreadsheets are used for cannot be solved in Kale. We describe an evaluation of the impact of Kale's restrictions by manually porting 60 randomly-selected spreadsheets from the EUSES spreadsheet corpus~\cite{Fisher2005:EUSES}. We successfully translated all of the spreadsheets to use Kale's reference approach. Nonetheless, due to the restrictions, we regard Kale as a design exploration that shows the potential benefits of controlling unsafe references; it is possible that future systems that relax some of these restrictions could be more usable in practice.

This paper describes the following contributions:

\begin{enumerate}
    \item A new spreadsheet system, Kale, which prevents users from introducing the aforementioned bugs with new syntax and semantics for spreadsheet reference (\Cref{sec:Kale}).
    \item A user study (\Cref{sec:user-study}) showing that when users of a traditional spreadsheet, Google Sheets, re-structure documents, they often introduce latent spreadsheet bugs. In contrast, users are more often able to complete those tasks successfully using Kale.
    \item A corpus study showing that most existing spreadsheets can be represented in Kale (\Cref{sec:corpus-study}).
\end{enumerate}

\section{The Kale Spreadsheet System}
\label{sec:Kale}

\subsection{Design}
The design of Kale prevents errors that arise from reference instabilities caused by structural changes to spreadsheets. Unlike traditional spreadsheets, Kale does not allow references to arbitrary rectangular ranges. Instead, references can refer to individual cells, whole columns, or whole rows.

Unlike traditional spreadsheet systems, which have extremely large grids, Kale sheets are organized into \emph{tables}, akin to tables in Apple's Numbers spreadsheet system. In the rest of this paper, the word \emph{table} will be used to refer to a grid of cells, regardless of whether that grid has configurable (like Numbers) or extremely large (like Sheets or Excel) extent. In Kale, tables have a header row that defines names of columns. Row indices start at 0, and row 0 is the row \emph{below} the header row. The header row cannot be referenced. \Cref{fig:t3kale} shows an example from the Payroll task in our evaluation.

Kale supports the following reference types; examples assume that there is a column called \texttt{Col} that can be referenced.

\begin{itemize}
    \item Single-cell references refer to the content in the referenced cell and are updated to refer to new locations when the cell is moved. These can be absolute (e.g., \texttt{Col[0]} refers to the 0th index in the Col column) or relative (e.g., \texttt{Col[+1]} refers to the cell in the Col column that is one row below the referencing formula). 
    \item Column references refer to all non-header cells in the referenced column. For example, \texttt{SUM(Col)} sums all non-header cells in the Col column.
    \item Row references refer to all cells in the referenced row: \texttt{SUM([0])} sums all cells in the row at index 0.
\end{itemize}

Kale's new definition of \emph{absolute} and \emph{relative} enables users to encode their intent when writing a formula. If a formula refers to a row that is moved (because of a row insertion, deletion, a sort operation, or a drag/drop operation), the new referenced row depends on whether the reference is absolute or relative. If a reference is absolute (e.g., \texttt{Col[0]}), and the referenced row is moved, then afterward, the reference points to the new location of the moved row. If relative (e.g., \texttt{Col[+1]}), the reference then points to the same relative offset as before, referring to different data. Absolute references are represented in terms of unique IDs, which Kale assigns to each row of each table. As a result, when rows are inserted, deleted, or permuted, formulas that use absolute references continue to refer to the same rows as they did before the changes.

Kale's formula syntax is traditional in that it supports infix arithmetic operators and function calls with parenthesized lists of comma-separated arguments. Because of the novelty of the reference syntax and semantics, we show the reference syntax in \Cref{fig:refSyntax}. The reference syntax is designed to maximize expressiveness. In addition to supporting relative and absolute row offsets that are expressed as literal numbers, Kale also supports relative offsets that are arbitrary expressions. For example, \texttt{Col[+1+1]} is a relative reference to the column called \texttt{Col} two rows below the current row. This addresses similar needs as those addressed by the traditional \texttt{OFFSET} function, which offsets a computed number of rows and columns from a given cell coordinate. Computed absolute indices are not supported because when users specify absolute indices, Kale must resolve them to row IDs at parse time.

\begin{figure}[tb]
    \centering
\begin{tabular}{p{3cm} r l l}
relativeRowRef & \bnfdef &  \textquotesingle [\textquotesingle{} \textquotesingle -\textquotesingle{} expr \textquotesingle ]\textquotesingle \\
	& \bnfalt  & \textquotesingle[\textquotesingle{} \textquotesingle +\textquotesingle{} expr \textquotesingle]\textquotesingle{} \\

absoluteRowRef & \bnfdef & \textquotesingle[\textquotesingle{} INT \textquotesingle]\textquotesingle{} \\

relativeCellOrColRef & \bnfdef & IDENT \\
	& \bnfalt & IDENT \textquotesingle [\textquotesingle{} - expr \textquotesingle]\textquotesingle{} \\
	& \bnfalt & IDENT \textquotesingle [\textquotesingle{} + expr \textquotesingle]\textquotesingle{} \\
 
absoluteCellOrColRef & \bnfdef & IDENT \textquotesingle [\textquotesingle{} INT \textquotesingle]\textquotesingle{}\\
\end{tabular}
    \caption{Kale reference syntax. Kale uses the ANTLR parser generator~\cite{ANTLR}, which uses the first matching rule in cases of ambiguity. Brackets, when present, always signify row indices or offsets. Note how + and - are used to specify that an index is relative rather than absolute. Literal indices are required for absolute references; other expressions that evaluate to numbers have relative reference semantics.}
    \label{fig:refSyntax}
\end{figure}

\begin{figure}
    \centering
    \begin{subfigure}{.5\linewidth}
    \centering
        \includegraphics[width=.4\linewidth]{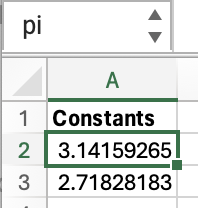}
        \caption{Named cells in Excel}
    \end{subfigure}%
    \begin{subfigure}{.5\linewidth}
    \centering
        \includegraphics[width=.7\linewidth]{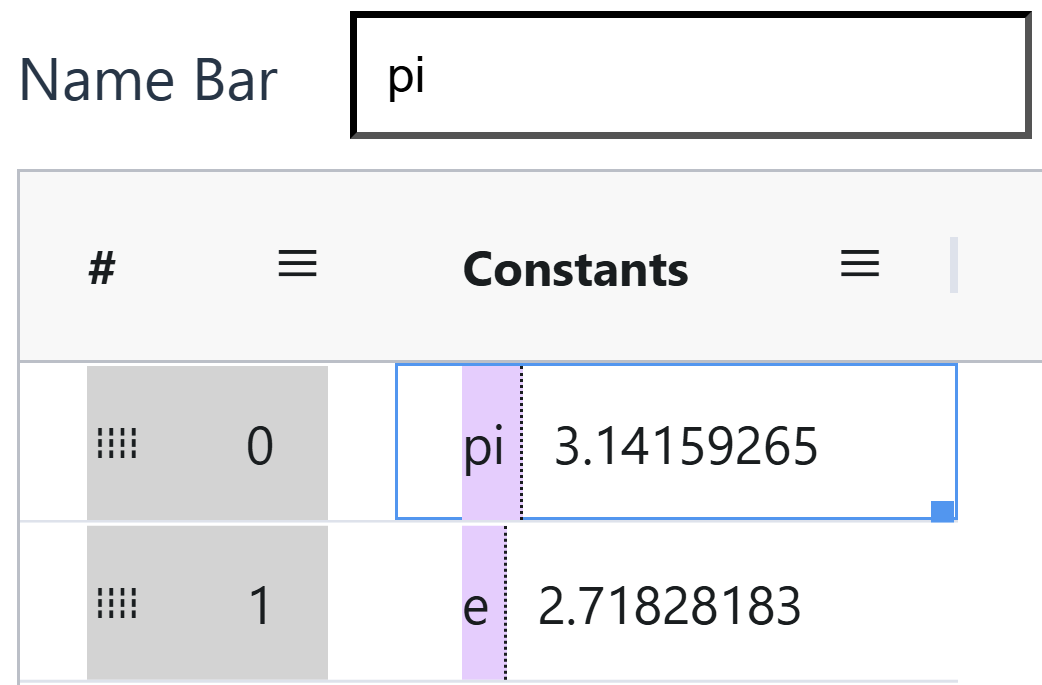}
        \caption{Named cells in Kale}
    \end{subfigure}
    
    \caption{Excel only shows one name at a time. Kale shows names in each named cell.}
    \label{fig:names}
\end{figure}

Some spreadsheet systems (including Excel and Google Sheets) enable users to use a dedicated interface to give \emph{names} to specific regions. After doing so, formulas can refer to those ranges by name. For example, \texttt{SUM(Sales)} refers to a region called \texttt{Sales} that was previously bound to a particular range. To enable name-based reference, Kale provides a \emph{name bar}, a text field that enables users to name individual cells. When the user wants to reference a cell in a formula, they can use either the location of the cell in the grid or the name of the cell. Like Excel and Sheets, names move with cell contents on cut/paste operations. \emph{Unlike} Excel and Sheets, names also move during sort operations. In Excel and Sheets, cell names are only visible when the cell is selected, and only in a dedicated name affordance. In Kale, every named cell always shows its name so that users can easily use them when writing formulas. \Cref{fig:names} compares the name user interfaces in Excel and Kale.

\subsection{Implementation}
Kale is a web application implemented with Typescript and React. Kale uses an off-the-shelf HTML/CSS/JavaScript table implementation, AG Grid~\cite{AGGrid}. Kale uses AG Grid’s core spreadsheet features, which include columns, rows, cells, filtering, selection, and editing. We also implemented some original features, including naming cells, representing and evaluating formulas, dependency graphing, and drag-drop.

\section{Preliminary Study of User Expectations}
\label{sec:preliminary-study}

\subsection{Methods}
To understand users' expectations regarding traditional spreadsheet behavior, we first conducted an IRB-approved survey-based preliminary study asking five participants to predict the results of several spreadsheet operations. We recruited participants using posters and our own contacts. All five participants were graduate students at a public R1 university, from fields of study including computer science, data science, and business analytics. In each question, we provided participants with a scenario and several multiple-choice options depicting possible results. The first four questions asked about the behavior of traditional absolute references (as implemented in Excel). The remaining three questions asked about how formulas are updated in response to structural table changes. Survey questions and results are displayed in \Cref{tab:survey}.

\begin{table*}[]
    \centering
    \caption{Summary of survey questions and results from the preliminary study.}
    \begin{tabularx}{\columnwidth}{@{}lp{0.3\linewidth}llX@{}}
    \toprule
     & Topic & Right & Wrong & Participant misconceptions \\
    \midrule
    Q1 & We asked how relative references work when using drag-fill. & 4 & 1 & Expected absolute column and relative row reference. \\
    \midrule
    Q2 & Drag-fill for \texttt{\$A2}. & 2 & 3 & Expected behavior of \texttt{\$A\$2}. \\
    \midrule
    Q3 & Drag-fill for \texttt{A\$2}. & 4 & 1 & Expected behavior of \texttt{\$A2}. \\
    \midrule
    Q4 & Drag-fill for \texttt{\$A\$2}. & 4 & 1 & Expected behavior of \texttt{\$A2}. \\
    \midrule
    Q5 & Behavior of row insertion directly above a range, as shown in \Cref{fig:Q5}. & 0 & 5 & Expected that the new row would be added to the range, or that the range would remain the same. In reality, the new row is not included and the range shifts down one row. \\
    \midrule
    Q6 & Behavior of row insertion in the middle of a range. & 3 & 2 & Expected that the range reference would remain the same. In reality, the new row is included in the range. \\
    \midrule
    Q7 & Behavior of row insertion directly below a range. & 4 & 1 & Expected that the range reference would include the new row. In reality, the new row is not included. \\
    \bottomrule
    \end{tabularx}%
    \label{tab:survey}
\end{table*}

\begin{figure}
    \centering
    \includegraphics[width=\linewidth]{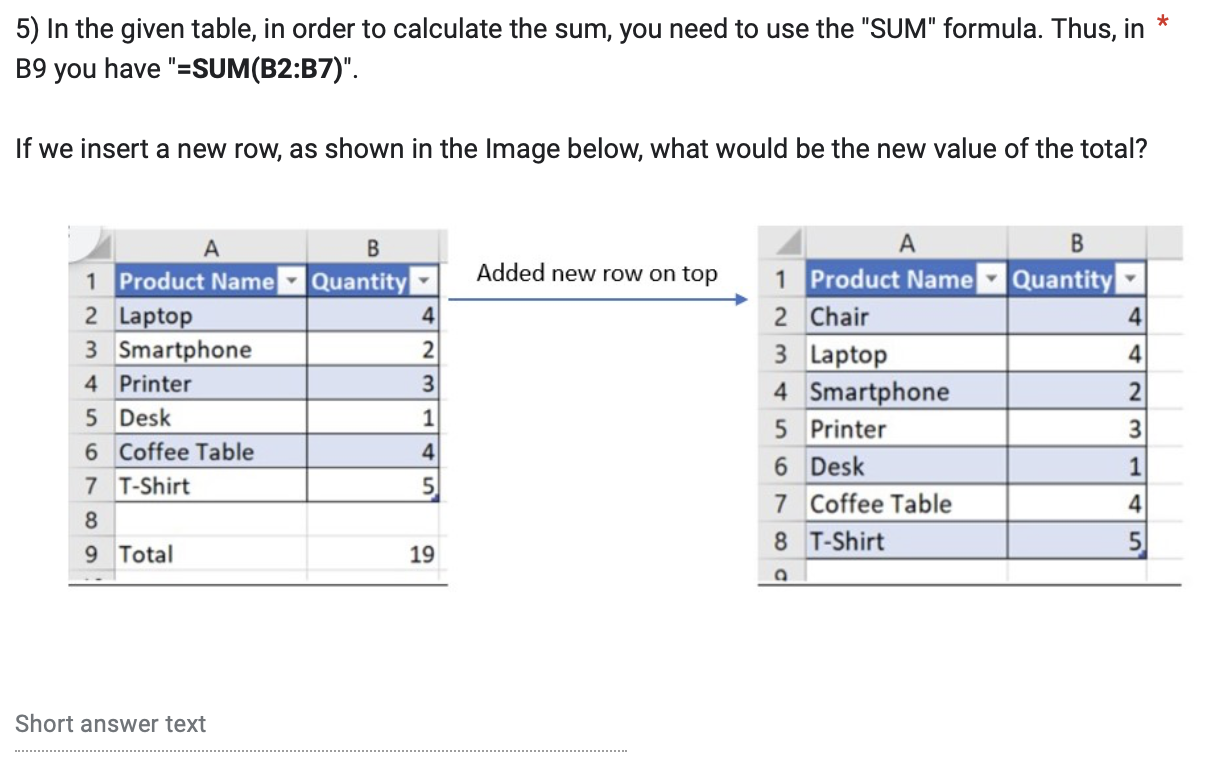}
    \caption{Question five regarding traditional spreadsheet behavior.}
    \label{fig:Q5}
\end{figure}

\subsection{Results and Discussion}
One participant answered all four questions about absolute references correctly; three got 3/4 correct; one got 2/4 correct. No one correctly answered all of the questions about how references are updated when rows are inserted; two answered 2/3 correctly and the remaining three answered 1/3 correctly.

Most participants had a correct understanding of traditional spreadsheet absolute references. All but one participant responded to at least three of the four questions correctly. However, for the questions about reference transformations on row insertion, there were only seven correct answers out of 15 responses, for a success rate of about 47\%. Although it is possible that more expert spreadsheet users might know the answers to these questions, spreadsheet systems are targeted at naive users, not only at experts; we view the semantics of references as fundamental to basic spreadsheet usage. The latent relationship between references and referenced cells means that users who insert rows may be unaware of which formulas might reference those rows. Combined, these two facts suggest that (a) when users write formulas, they write them without regard for how they will be updated when table structures change; (b) when users make structural changes to tables, they cannot conveniently assess whether their changes have broken any referencing formulas.

\section{User Study}
\label{sec:user-study}

\subsection{Methods}
\label{sec:overview}
Given our observation in \cref{sec:preliminary-study} that users may be unaware of the effects of structural changes on formulas, we turned to assess the risk of bugs resulting from these misunderstandings and evaluate Kale's efficacy as a solution.

Specifically, we designed a user study aiming to explore the following research questions:
\begin{itemize}
    \item \textbf{RQ1}: To what extent do the risks described in \Cref{sec:intro} cause bugs in tasks using traditional spreadsheets?
    \item \textbf{RQ2}: Is there a difference in the risk of encountering these bugs with Kale, as compared to traditional spreadsheets?
    \item \textbf{RQ3}: Is there a difference in overall task correctness with Kale, as compared to traditional spreadsheets? Unlike RQ2, this considers errors beyond those caused by the risks that we have defined.
    \item \textbf{RQ4}: Is there a difference in task completion speed with Kale, as compared to traditional spreadsheets?
\end{itemize}

We conducted a between-subjects, IRB-approved study with 25 participants, who completed spreadsheet tasks using either Google Sheets or Kale. Participants were recruited via postings on online student forums associated with two public R1 universities, as well as our own contacts and snowball sampling (in which we asked participants to refer their contacts). No participants from the preliminary study were included in this sample. During recruitment, we obtained informed consent from participants and asked them to complete a survey asking about demographic information, including age, gender, occupation, educational background, field of study, and self-estimated experience and frequency of usage with spreadsheets. All user study sessions were conducted remotely via Zoom. The study took an hour for each participant. We compensated each participant with a \$20 gift card.


In order to obtain a balanced sample with approximately equal numbers of participants in both of our two conditions, we manually assigned participants to conditions based on their self-estimated experience with spreadsheets and, as many of our participants were students, field of study. We assigned 12 participants to use Sheets and 13 to use Kale.

We provided participants with a document providing the details of each task and links to spreadsheets, one for each task, in Google Sheets or a web-based implementation of Kale. We selected Sheets for our traditional spreadsheet condition since it is available as a web application, allowing us to share task spreadsheets with participants via links without requiring them to download software. Participants assigned to use Sheets were permitted to download the spreadsheet from Sheets and use any spreadsheet tool of their choice (e.g. Excel or Numbers); however, all 12 participants assigned to this condition elected to use Sheets.

Participants assigned to use Kale were shown a tutorial video (approx. 6 minutes long), and additionally had access to a Kale help document, which they could refer to for assistance with syntax and other tool-specific functionalities. Participants were allowed access to internet tools (e.g. Google search) during the study. This was intended to enable participants assigned to Google Sheets to look for documentation and help online. However, we disallowed participants from using generative AI tools (e.g. ChatGPT) to prevent them from generating formulas, which we felt could disrupt the formula-writing process in which we are interested.

While participants completed tasks, we recorded their screens and audio. We also collected the final spreadsheets created by the participants. This allowed us to analyze the end results of their tasks, focusing on occurrence of errors, and to note different formulas used by participants to complete the tasks. There were no per-task time limits. However, the overall study duration was limited to an hour.



\subsection{Participants}
We recruited 25 participants; 12 identified as female, 11 as male, one as non-binary, and one declined to state. Their average age was approximately 23, with a range from 18 to 27. By their self-described occupations, 11 participants were students. The remaining participants were employed in diverse occupations, including project management, software engineering, music teaching, and cinema programming. Participants also studied in or had graduated with degrees in a diverse set of fields, most common among them data science (6), cognitive science (5), and economics (5). While most participants majored in STEM fields, there were also participants who studied film, English literature, and music.

Participants also reported their self-estimated experience with spreadsheets, rated from one (least experienced) to ten (most experienced). The mean was approximately 5.44, with a minimum of two and a maximum of nine. This metric was used to assign participants to balanced groups; the mean self-estimated experience was 5.50 in the Sheets condition, out of 12 participants, and 5.38 in the Kale condition, out of 13 participants. Field of study was also used to inform balancing; participants were assigned such that, between conditions, there was a balance of participants from different fields, as well as broader areas like sciences, engineering, and humanities.

\subsection{Tasks} 

We designed four tasks, T1, T2, T3, and T4, to evaluate the risk of users introducing bugs in their spreadsheet via R1, R2, R3, R4, and R5. To improve their real-world plausibility, spreadsheets modified during these tasks were based off of real spreadsheets from the EUSES spreadsheet corpus~\cite{Fisher2005:EUSES}. Each task was also coupled with a real-world scenario involving modifying the spreadsheet and sharing it with a fictional coworker or colleague; this scenario was intended to encourage participants to ensure their formulas were correct, even after structural transformations, without explicitly requiring them to double-check the formulas. Tasks were divided into two to six parts, which participants were instructed to complete in order.

Participants in both conditions were given the same tasks. Some minor changes were required to adapt the spreadsheet tasks to Kale; most notably, T1, T2, and T3 asked spreadsheet users to fill in a table of summary statistics, while Kale users were instead asked to fill in the last row of the sheet, due to Kale's requirement that all sheets be structured as a single table. Since Kale disallows spaces in column names, some columns were given minor naming changes. One part of T3 also asked participants to bold a row; this instruction was omitted from the Kale version of this task, as bolding text is not implemented in Kale.

We designed each task to enable evaluation of specific reference transformation risks as follows:

\begin{figure*}[tb]
    \begin{subfigure}{\columnwidth}
        \centering
        \includegraphics[width=\linewidth]{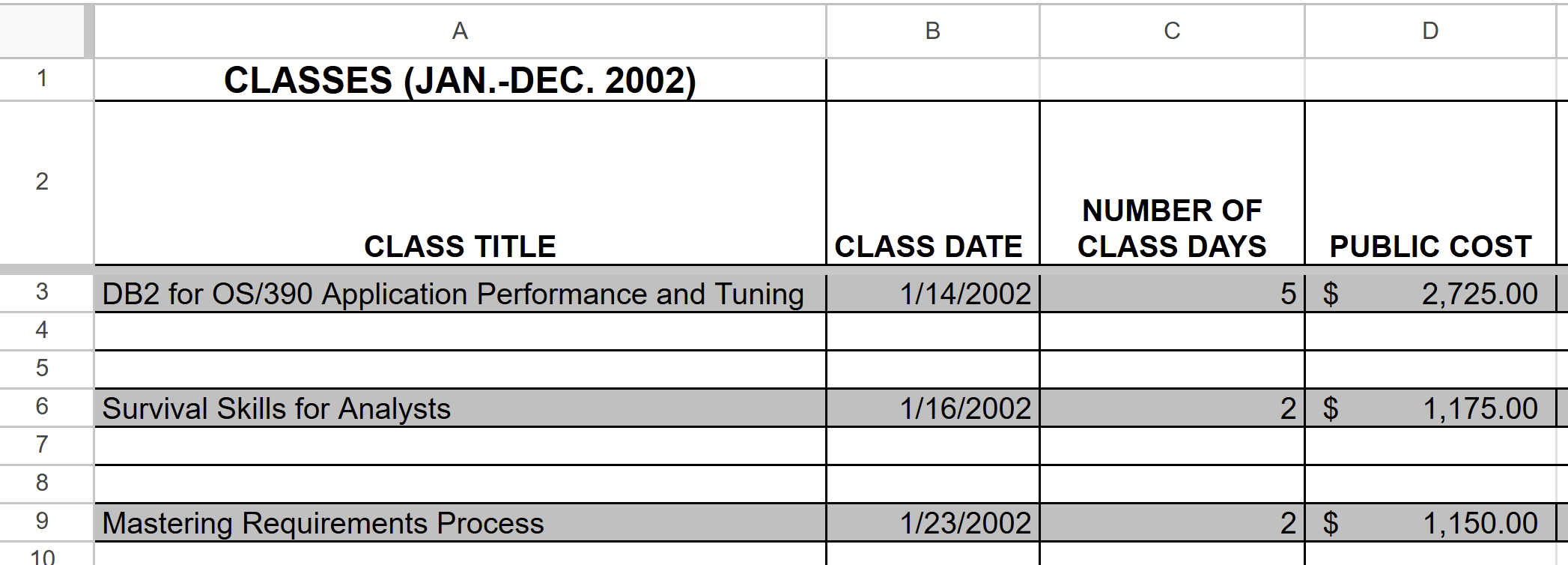}
        \label{fig:t1sheets}
    \end{subfigure}
    \begin{subfigure}{\columnwidth}
        \centering
        \includegraphics[width=\linewidth]{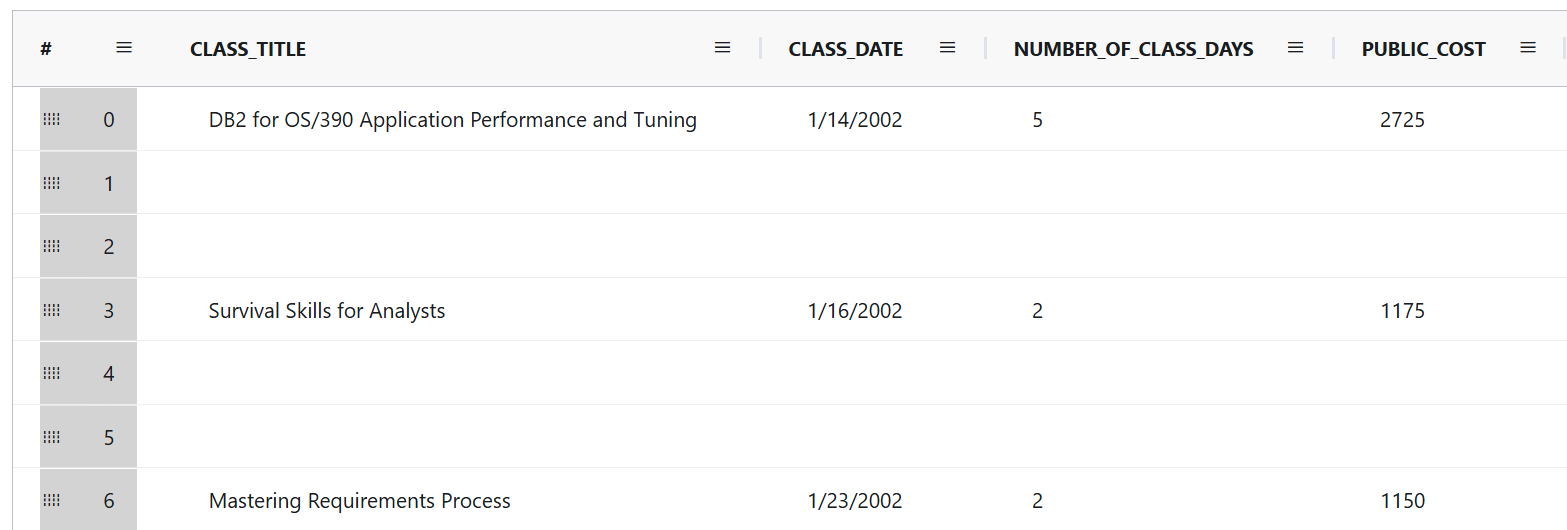}
        \label{fig:t1kale}
    \end{subfigure}
    \caption{The Classes task (T1), shown in Sheets on the top and Kale on the bottom.}
    \label{fig:t1}
\end{figure*}

\textbf{T1: Classes (risk R1)}: We provided participants with a spreadsheet of classes for a professional development company, with each entry in the table consisting of three rows.\footnote{The spreadsheet used in T1 is based closely on the spreadsheet \texttt{02rise.xls} from the \texttt{financial} section of the EUSES corpus. Our version is significantly scaled down, with many rows and columns removed.} \Cref{fig:t1} shows this task in Google Sheets and Kale. In Part 1 of this task, we asked participants to use formulas to compute summary statistics across columns of the table, including the total of one column (``NUMBER OF CLASS DAYS''), and the averages of two columns (``PUBLIC COST'' and ``COMPANY COST''). In Part 2, we then asked participants to add new data to the table, consisting of three new entries, from another sheet, and to insert the entries into the table such that it remained sorted chronologically. Since the correct placement of two of these entries involves adding new rows to the beginning and end of the table, this transformation causes formulas in traditional spreadsheets referencing only the existing range of rows to break, incurring risk R1. 
    
\begin{figure*}[tb]
    \centering
    \resizebox{0.85\linewidth}{!}{%
    \includegraphics[width=\linewidth]{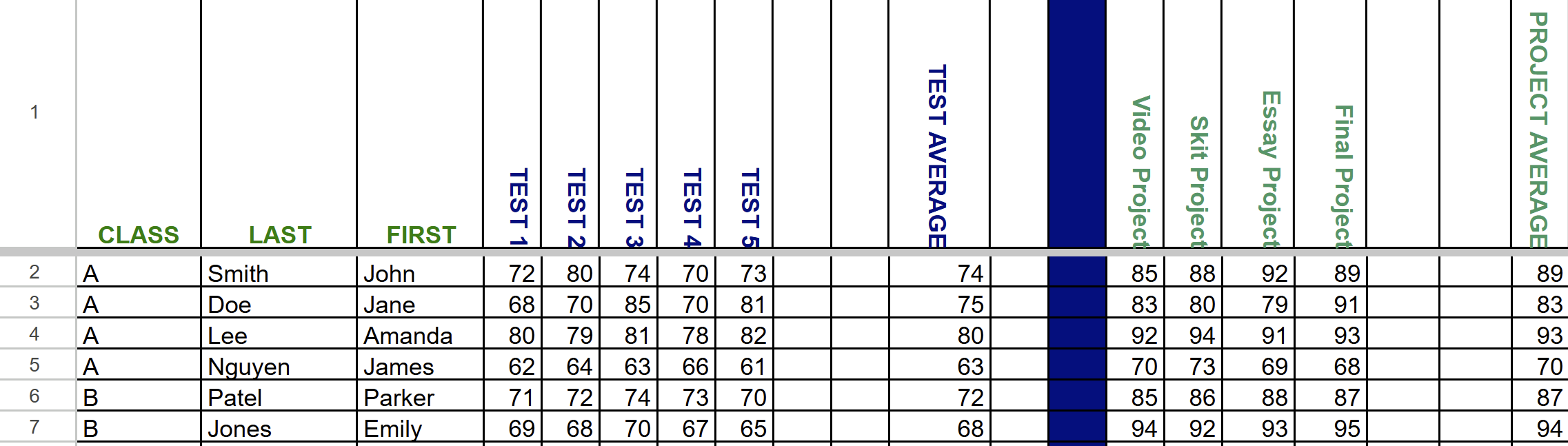}%
    }
    \caption{The Gradebook task (T2), shown in Sheets.}
    \label{fig:t2sheets}
\end{figure*}

\textbf{T2: Gradebook (risks R1, R2, and R4)}: We tasked participants with modifying a mock teacher's gradebook with test and project scores for students in different classes.\footnote{The spreadsheet in T2 is based on a sheet in \texttt{7\_gradebook\_xls.xlsx} from the \texttt{grades} section of the EUSES corpus. Compared to the original, we removed some rows and columns, and filled the table with mock data.} \Cref{fig:t2sheets} displays this task in Google Sheets. This was the only task to involve multiple risks; across the formulas and structural transformations involved in this task, we identified four potential risks: two instances of R1, and one of R2 and R4. In Part 1, we asked participants to add a new ``Intro Project'' column, located in another sheet, into the table. The original (unmodified) spreadsheet includes a column titled ``Project Average'' taking the average across columns in the Projects category. This formula is written in Sheets using a range, and in Kale using an average of individual cells. In both conditions, the new column ``Intro Project'' added in Part 1 is not automatically included in the formula, incurring risk R1. In Part 2, participants wrote a formula to compute the average test score across all students, as well as a formula in Part 3 to compute the average test score of only Class A. In Part 4, participants added new rows corresponding to students from a new class, Class E, to the table. The rows added to the table during this part may not be included in the average of all student test scores from Part 1 if the formula is written using a range across only the existing rows. This incurs risk R1 for a second time. In Part 5, participants then computed the average of three named cells (named ranges in Sheets) holding scores of ``fan favorite'' projects. Finally, in Part 6 participants sorted the table in alphabetical order by student last name, permuting the rows. If the formula from Part 3 computing the average test score of students in Class A was written using a reference to the rectangular range of cells containing the scores of students in Class A, sorting caused this formula to reference different data, incurring risk R2. In addition, if named ranges were used in Sheets to compute the average score of fan favorite projects in Part 5, sorting also produced another risk; in Sheets, named ranges do not move when rows are sorted, leading to risk R4.

\textbf{T3: Payroll (risk R3)}: We provided participants with a table of pay rates at an IT company for different roles, across four time periods.\footnote{The spreadsheet in T3 is nearly identical to a sheet in \texttt{IT\_Schedule\_Rates.xls} from the \texttt{database} section of the EUSES corpus. Compared to the original, we removed two hidden columns.}  \Cref{fig:t3kale} shows this task in Kale. In Part 1, we asked participants to write a formula computing the average pay across four different ``software engineering'' roles in the most recent time period (the rightmost column). In Part 2, we asked participants to swap the values in two rows, one of which corresponded to the bottommost software engineering role. Finally, in Part 3 we asked participants to move the row for the topmost software engineering role up to the first row of the table, and, in the Sheets condition, bold the row. If the average pay across software engineering roles is computed using a range of cells, this task involves moving the first and last row in that range. If these transformations are performed in Sheets via drag/drop or cut/paste, the range expands to incorrectly include the cells between, incurring risk R3.

    
\textbf{T4: Aquarium (risk R5)}: We asked participants to fill in an incomplete spreadsheet predicting the populations of fish in different tanks at an aquarium over a period of five months, with one column for each month\footnote{The spreadsheet in T4 is loosely inspired by \texttt{CANTILEV.xls} from the \texttt{filby} section of the EUSES corpus. The use case for and data from this spreadsheet are entirely different from ours; however, we attempted to replicate the original spreadsheet's structure and layout (e.g. its practice of storing constants in a separate table).}. In Part 1, participants first used formulas to fill two columns with the values in the previous (left-adjacent) column, multiplied by a growth rate and rounded to the nearest integer. We provided participants with a named cell in the sheet holding the value of the growth rate, as well as a formula to reuse and adapt: \texttt{=ROUND(n * growth\_rate, 0)}. In Part 2, we asked participants to fill in the remaining two columns according to the same pattern, but with an added flat value changing between the two months. The value to add each month was stored in a separate table in the Sheets condition, and in additional columns in the Kale condition. Unlike previous tasks, T4 was designed to assess risks associated with confusion between relative and absolute references; if the flat value to add each month is accessed using a relative reference, then drag-filling or copy-pasting the formula will not correctly fill in the rest of the column, leading to risk R5.

\begin{figure}[tb]
    \centering
    \includegraphics[width=\linewidth]{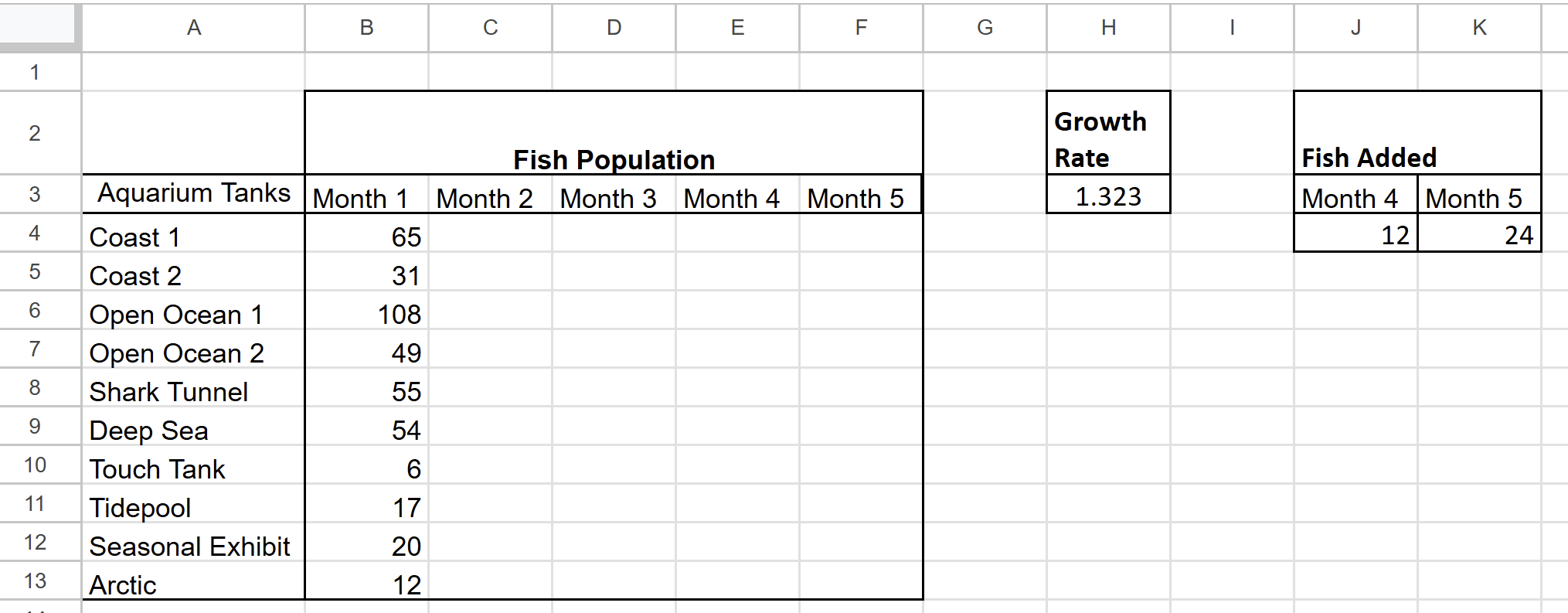}
    \caption{The Aquarium task (T4), shown in Sheets.}
    \label{fig:t4sheets}
\end{figure}


\subsection{Results}

To answer our research questions, we collected participants' final spreadsheets. Three participants ``timed out'' and did not complete all tasks before the end of the study duration. The resulting incomplete or unattempted tasks are excluded from all analyses, as well as one task where a participant using Sheets was, due to an error, sent a link to a spreadsheet for task T4 that had already been completed. One participant using Sheets timed out after completing three tasks, while two participants using Kale both timed out after completing two tasks.

\begin{figure*}[tb]
    \centering
    \includegraphics[width=\linewidth]{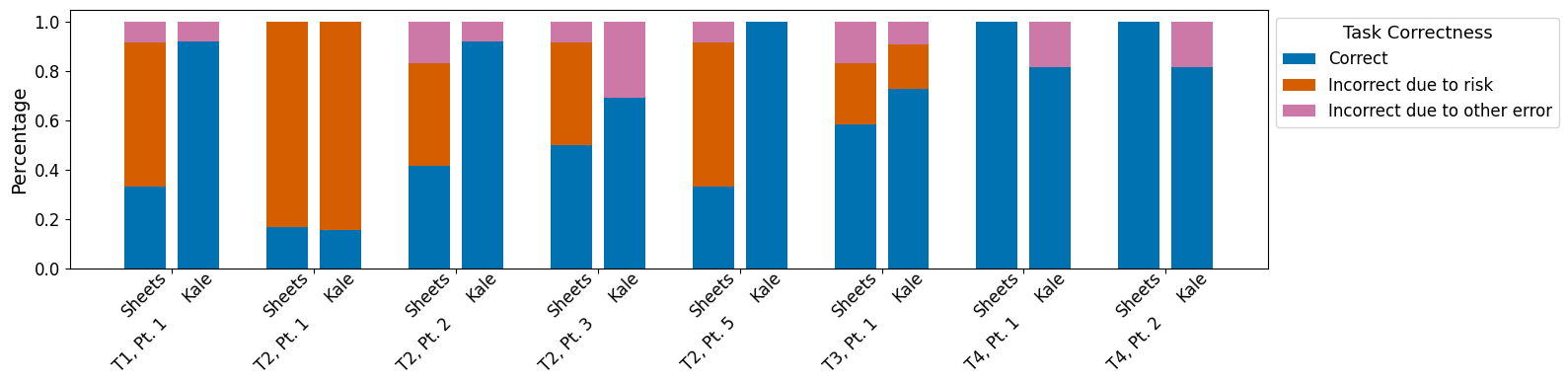}
    \caption{Breakdown of participant correctness percentages for formula-related task parts, by condition and task part.}
    \label{fig:formula-correctness}
\end{figure*}

To assess task correctness, we manually compared participants' spreadsheets to a reference solution. We measured correctness separately for each part of a given task. Since we are interested primarily in formula errors, we limit our analysis to parts of tasks involving writing or editing formulas. Across all four tasks, there were eight such formula-editing parts. For incorrect formulas, we recorded whether they were caused by a risk or some other error. For incorrect formulas exhibiting both a risk and another error, we considered them incorrect due to the risk. \Cref{fig:formula-correctness} displays the percentage of participants in each part and condition that wrote correct formulas, wrote incorrect formulas due to risks, and wrote incorrect formulas due to a different error. Kale exhibited a higher correctness rate in five task parts, as well as a lower or equal rate of non-risk-related errors in another five parts. We conducted a Wilcoxon exact test measuring the effect of condition on the total number of correct formula tasks, finding that the difference in total correctness is not statistically significant ($p = .025$)\footnote{As described later, we conduct seven other hypothesis tests on risk-related error rate. Since risk-related error rate is not independent of task correctness, we perform a Bonferroni correction on these eight hypothesis tests to control the probability of a type I error. Our new significance level for these hypothesis tests only is $\alpha = 0.05/8 = 0.00625$.}.

\begin{figure}[tb]
    \centering
    \includegraphics[width=\linewidth]{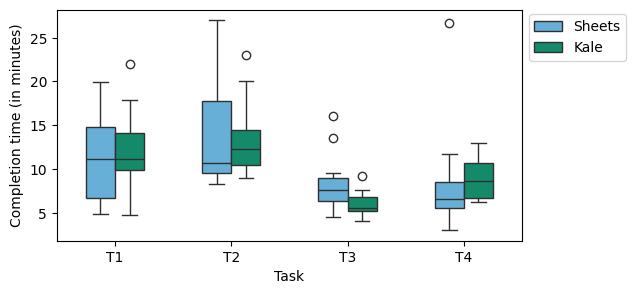}
    \caption{Distributions of task completion time in minutes by task and condition. \textit{Lower is better.}}
    \label{fig:task-timing}
\end{figure}

To assess task completion times, we recorded start and end times for each task. \Cref{fig:task-timing} displays the distributions of task completion times in each condition. According to a Wilcoxon exact test, Kale users experienced a statistically significant speedup in Task 3 ($p = .049$), but differences in timing were not significant in the other tasks ($p = .69$, $p = .85$, and $p = .15$ for Tasks 1, 2, and 4 respectively).

To assess the occurrence of risks, we manually inspected participants' spreadsheets and assigned occurrences of each risk to one of five categories, abbreviated A, C, I, N, and E:

\begin{itemize}
    \item \textbf{Risk avoided (A)}: The participant avoided this risk by writing a formula that prevented the bug from occurring.
    \item \textbf{Risk fixed correctly (C)}: The participant encountered the associated bug, but noticed and correctly fixed it.
    \item \textbf{Risk fixed incorrectly (I)}: The participant encountered the associated bug and noticed it, editing the formula, but did not correctly fix the bug.
    \item \textbf{Risk not fixed (N)}: The participant encountered the associated bug and did not notice it, or otherwise did nothing to edit the formula.
    \item \textbf{Risk not encountered due to error (E)}: The participant inserted an error elsewhere during the task that prevented them from encountering the risk.
\end{itemize}

The counts of participants in each category for each task-risk combination is shown in \Cref{tab:combined-risk-results}. For participants who encountered risks, we also noted the ways in which they caused them; for those who avoided risks, ways in which they wrote formulas to avoid them; and for participants who incorrectly fixed risks, ways in which the fixes were incorrect.



\begin{figure}[tb]
    \centering
    \includegraphics[width=\linewidth]{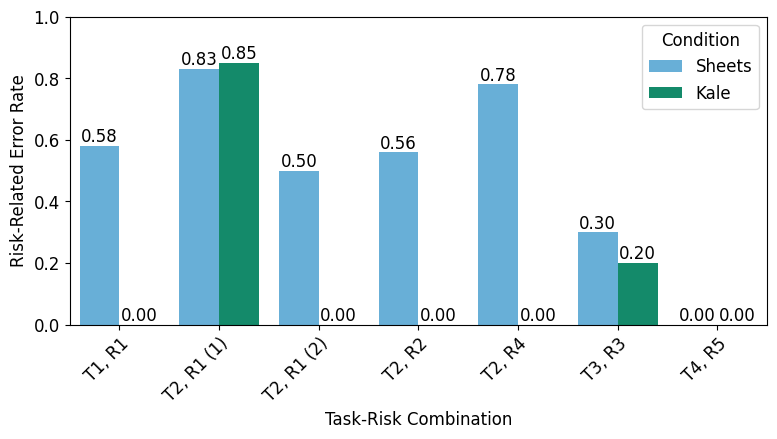}
    \caption{Rates of risk-related error by condition for each task-risk combination. \textit{Lower is better.}}
    \label{fig:risk-related-error}
\end{figure}

We additionally computed the rate at which participants in each condition ultimately produced errors related to each risk. We grouped categories I and N together to produce a count of participants who submitted formulas with errors due to the corresponding risks, and grouped A and C together to count participants who did not produce risk-related errors. We conducted hypothesis tests using these counts (comparing \(I + N\) against \(A + C\)). We additionally calculated the rate of risk-related errors for each task-risk combination as \(\frac{I + N}{A + C + I + N}\). That is, we divided the number of participants who encountered the risk and did not correctly fix it by the total number of participants, excluding \(E\), the group of participants who could not have encountered the risk due to other errors. \Cref{fig:risk-related-error} shows this risk-related error rate for each task-risk combination in both conditions. We note several observations, which we organize by risk.

\begin{table}[]
    \centering
    \caption{Summary of risk categories in both conditions. A: avoid; C: fixed correctly; I: fixed incorrectly; N: not fixed; E: not encountered due to error}
    \label{tab:combined-risk-results}
    \begin{tabular}{llrrrrr|rrrrr|r}
    \multirow{2}{*}{} & \multirow{2}{*}{} 
        & \multicolumn{5}{c}{Sheets} 
        & \multicolumn{5}{c}{Kale} \\
    \toprule
    \textbf{Task} & \textbf{Risk} & \textbf{A} & \textbf{C} & \textbf{I} & \textbf{N} & \textbf{E} & \textbf{A} & \textbf{C} & \textbf{I} & \textbf{N} & \textbf{E} & \textbf{$p$-value} \\
    \midrule
    T1 & R1     & 1  & 4 & 4 & 3  & 0 & 13 & 0 & 0 & 0 & 0 & \textless.001 \\
    T2 & R1 (1) & 0  & 2 & 0 & 10 & 0 & 0  & 2 & 0 & 11 & 0 & .93 \\
    T2 & R1 (2) & 2  & 3 & 0 & 5  & 2 & 13 & 0 & 0 & 0  & 0 & .001 \\
    T2 & R2     & 0  & 4 & 0 & 5  & 3 & 12 & 0 & 0 & 0  & 1 & .001 \\
    T2 & R4     & 2  & 0 & 0 & 7  & 3 & 13 & 0 & 0 & 0  & 0 & \textless.001 \\
    T3 & R3     & 0  & 7 & 1 & 2  & 2 & 8  & 0 & 0 & 2  & 1 & .60 \\
    T4 & R5     & 11 & 0 & 0 & 0  & 0 & 11 & 0 & 0 & 0  & 0 & N/A \\
    \bottomrule
    \end{tabular}
\end{table}

\textbf{\ronefull}. In Sheets, all three instances of risk R1 were encountered by some participants. These participants' responses to this risk varied; in all three instances, some participants were able to identify and fix the error, but some others did not notice or otherwise attempt a fix. In T1, four participants attempted to fix the error but only edited \emph{one} end of a range that had extended in both directions. One participant in T1 and two in T2 were able to avoid instances of this risk using Sheets by referencing whole columns instead of ranges of cells (e.g. referencing the whole column A with range \texttt{A:A}). In Kale, all 13 participants were able to entirely avoid the first and third instances of risk R1 by writing formulas referencing whole columns. However, the second instance of this risk, a formula averaging only part of each row, went undetected by most participants, and, unlike the other cases of risk R1, was not automatically fixed by whole-column references in either condition. Kale fully eliminated errors caused by the first and third instances of R1, but in the second instance, the rate of risk-related errors was slightly higher with Kale. A $\chi^2$ test shows that Kale yielded a statistically significant decrease in the rate of risk-related errors in the first and third instances ($\chi^2 (1, N = 25) = 6.67, p = .0003$ and $\chi^2 (1, N = 23) = 10.22, p = .0014$, respectively), but finds no significant relationship in the second ($\chi^2 (1, N = 25) = 0.01, p = .93$).

\textbf{\rtwofull}. In Sheets, risk R2 caused five participants to submit errors, compared to four who were able to fix the error. Three participants did not encounter the risk, due to sorting the table incorrectly. No Sheets participants wrote formulas avoiding this risk. However, with the exception of one participant who skipped the sorting step, all 12 other Kale participants were able to avoid the risk using Kale's adjustment of references to sorted data. A $\chi^2$ test shows that Kale yielded a statistically significant decrease in the rate of errors related to this risk ($\chi^2 (1, N = 21) = 10.69, p = .0011$).

\textbf{\rthreefull}. In Sheets, seven participants were able to successfully identify and fix risk R3. Eight Kale users were able to fully avoid the risk using Kale's automatic adjustment of references in moved row. However, for two participants, this behavior introduced a new, related risk. These participants swapped the numerical row values in Part 2 by exchanging the job titles and moving the rows, rather than by copy-pasting the numerical values; Kale automatically updated references to these rows during the move, causing the referenced values to remain the same instead of being replaced. In categorizing this error, we decided to consider it an instance of R3, as it was generated by the same circumstances as R3 would be in Sheets. Despite this threat, a $\chi^2$ test finds no significant relationship between condition and rate of errors related to this risk ($\chi^2 (1, N = 20) = 0.27, p = .60$).

\textbf{\rfourfull}. In Sheets, the majority of participants (7/12) encountered R4, and once encountered no participant was able to identify and fix the error. Two participants were able to avoid the error by using the literal values of the named ranges in their formula, rather than referencing the named ranges themselves. In Kale, all 13 participants were able to avoid the error, since Kale automatically moves named cells during sorting. A $\chi^2$ test shows that Kale yielded a statistically significant decrease in the rate of errors related to this risk ($\chi^2 (1, N = 22) = 17.99, p < .0001$).

\textbf{\rfivefull}. In T4, across both conditions, all 22 participants attempting the task were able to avoid risk R5. Participants avoided this risk using a variety of methods, including hard-coding the added value instead of referencing a cell, using absolute references, or by defining new named cells. Since all participants had the same outcome, we do not perform a $\chi^2$ test for this risk.

\subsection{Discussion}

To answer RQ1 regarding the impact of our five risks, we refer to the rates of risk-related errors. To put these error rates into perspective, Panko~\cite{Panko2015:What} hypothesized that the average rate at which users make errors in spreadsheet cells is within the range of 1-5\%, with prior studies discovering an average error rate of 3.9\%. With the exception of R5, the rates at which participants using Sheets in our study inserted errors due to these risks (50-83\% for R1, 56\% for R2, 30\% for R3, and 78\% for R4) were orders of magnitude larger than Panko's estimated global cell error rate, suggesting that risks R1-R4 all present a distinct danger to spreadsheet users. In contrast, no participants encountered R5 (references are relative by default), suggesting either that this risk is not a danger to users, or that our task was ineffectively designed to assess the risk.

To address RQ2, we compare the incidence of risk-related errors between participants using Sheets and Kale. We found that, in four out of seven instances of risks, Kale completely eliminated risk-related errors that were present with rates of over 50\% in Sheets. Kale significantly reduces the occurrence of three out of five risks: R1, R2, and R4. Simultaneously, results for RQ3 and RQ4 show that participants do not significantly sacrifice correctness or speed by using Kale.

However, our results also reveal opportunities for tool improvement and further investigation. We found that Kale was not able to eliminate one variant of errors caused by R1: specifically, whole-column and whole-row references do not allow a range to automatically adjust to new data if that range does not span a whole table row or column. We also observed that, in a specific situation encountered by two participants in T3, Kale's sorting behavior actually introduced new errors. This may be because users' expectations have been shaped by their experience with traditional spreadsheets, which have subtly different sorting behavior. These issues partially explain Kale's failure to significantly prevent R3 in T3 and the first instance of R1 in T2, as reflected in both our hypothesis tests and the error rates shown in \Cref{fig:risk-related-error}.

\section{Corpus Study}
\label{sec:corpus-study}
\subsection{Methods}
Because Kale does not support arbitrary rectangular range references, it is possible that some spreadsheet tasks are difficult with Kale. If a formula needs to refer to only part of a row or column, then it might suffice to write the formula using multiple references. Otherwise, it might be necessary to restructure. 

To assess to what extent users might need to make major changes to their documents if they had used Kale to create them rather than a traditional spreadsheet, we conducted a corpus study. We began with the EUSES corpus~\cite{Fisher2005:EUSES}, which includes 4,233 Excel spreadsheets. After removing duplicates and those with minimal differences, we refined our dataset to 1,726 unique spreadsheets. We  categorized these into three groups based on their formulas:

\begin{itemize}
    \item Spreadsheets that included range references (e.g., \texttt{SUM(A1:B10)}): 677 spreadsheets (39\%).
    \item Spreadsheets with cross-sheet references (e.g., \texttt{Sheet1!C3}): 50 spreadsheets (3\%).
    \item Other spreadsheets (references were limited to single cells): 999 spreadsheets (58\%). 
\end{itemize}

Because of our interest in evaluating Kale, we sampled 50 spreadsheets from the first group, containing range references. We also sampled 10 spreadsheets with formulas but only single-cell references for comparison purposes. Then, three co-authors divided the documents among themselves and manually converted the documents to Kale, recording the time and changes required.

\subsection{Results and Discussion}
We were able to convert all 60 spreadsheets to use Kale. The time required ranged from 4 to 44 minutes, with a mean of 18 minutes and a median of 15 minutes. Spreadsheets with few range references were trivially easy to convert to Kale, while spreadsheets with many range references were sometimes challenging. Due to Kale's limited function library, we did not attempt to translate calls to unsupported functions.

This evaluation is limited because it did not include documents that use cross-table references. However, since those only comprise 3\% of the corpus, and the mean time was 18 minutes, we conclude that reference-related restrictions in Kale are unlikely to present significant barriers to real-world success.

\section{Related Work}

The flexibility and ease-of-use of spreadsheets have contributed to their global commercial success, but have also made them prone to errors~\cite{10.1145/3491102.3501833}. From a total of 85 intensive inspection studies, it was found that 94\% of the spreadsheets contained errors ~\cite{Panko2015:What}. In general, past studies have audited existing spreadsheets; our study observes participants using spreadsheets to see whether they identify errors as they occur.


Understanding the root causes of spreadsheet errors is crucial for addressing them. Numerous studies have explored common errors and their causes, considering both limitations of human cognitive processes~\cite{6480329} and specific kinds of errors. Panko~\cite{4755815} categorized spreadsheet errors into two types: blameless errors from innocent mistakes and culpable violations of laws or corporate practices. The causes of these errors are due to misinterpretation of domain knowledge, incorrect algorithmic expression, and implementation mistakes, which include slips and lapses during execution. 

In an effort to develop means of preventing and detecting errors, several other researchers have categorized spreadsheet errors. In a field study, Caulkins et al.~\cite{article} found that executives and managers frequently encounter inaccurate data, errors inherited from reuse of spreadsheets, model errors (including structural errors and errors of omission), and errors in the use of functions. Similarly, Badame et al.~\cite{6405299} explored spreadsheet refactoring and discovered that formulas often suffer from the same issues seen in professional software: hardcoded constants, duplicated expressions, unnecessary complexity, and unsanitized input. Barowy et al.~\cite{Barowy2018:ExcelLint} attributed some errors to the ease of use of drag-fill, which can result in correct references. Lastly, Reinhardt et al.~\cite{1357687} conducted a case study on computer literacy students, and found that their errors can be classified into either conceptual-related errors, or mathematical and logical related errors. None of these categories distinguish errors inserted via structural changes in spreadsheets from other kinds of errors. In this paper, we leverage this distinction to identify a new risk and possible solution space.


Several tools have also been developed to identify and quantify errors. Schalkwijk et al.~\cite{article1} proposed PerfectXL, a tool that visualizes spreadsheet dependencies and determines possible errors in spreadsheets based on previous findings. Barowy et al. ~\cite{10.1145/3276518} developed a static analysis tool, ExceLint, which uses the rectangular layout of spreadsheets to identify formula errors. Badame et al. ~\cite{6405299}. implemented REFBOOK, a plugin for Excel that implements seven refactorings that reliably remove spreadsheet smells, patterns or characteristics in spreadsheets that might indicate deeper problems. Rather than rectifying errors that appear in Excel, Kale provides users with a new platform for spreadsheet creation, manipulation, and error detection. Many related approaches to automated spreadsheet QA are discussed in Jannach et al.'s~\cite{Jannach2014} survey, including tools for formula visualization, static analysis, fault localization, testing, model-driven development, and design and maintenance of spreadsheets. Most similar to Kale are tools that provide enhanced safety guarantees in spreadsheets as they are being created, rather than testing, debugging, or analysis of existing spreadsheets.

Some approaches use features in existing spreadsheet tools; for example, Panko et al.~\cite{panko2008spreadsheet} proposed cell protection, or restricting modifications to pre-specified input cells, and re-keying data, a traditional data verification approach that reduces data input errors by requiring data to be entered twice in two separate input sections, allowing error detection by highlighting differences between two blocks of input data. 

Chalhoub et al.~\cite{10.1145/3491102.3501833} discovered that without formal types or data structures, spreadsheets suffer from classes of errors that in traditional programming languages are easily detected and prevented, and propose to introduce data structures that aim to strike a balance between the freedom and flexibility of the traditional grid, and the safety and power of formal types and structures. However, introducing these restrictions could restrict users' ability to flexibly use the cell grid. This insight is particularly relevant to Kale, as one potential drawback is that it limits users to structuring their spreadsheets as separate tables.

A number of tools have also, like Kale, been designed to prevent errors by enforcing structure on spreadsheet design. Many of these existing tools involve model-driven development: the use of a separate, abstract representation of a spreadsheet to ensure certain properties about its function, especially by generating provably correct formulas. Early work by Isakowitz et al. \cite{isakowitz_toward_1995} proposed the separation of spreadsheets into logical and physical layers representing the spreadsheet's functionality and implementation. Paine \cite{paine_ensuring_2008} introduced Model Master, a tool capable of compiling spreadsheets to and from a textual, declarative programming language. Erwig et al. \cite{erwig_automatic_2005, erwig_gencel_2006} introduced Gencel, a system for developing visual spreadsheet templates which can be used to automatically generate spreadsheets with defined update operations. Similarly to Kale, Gencel ensures that cell references stay valid under structural transformations. Engels and Erwig \cite{engels_classsheets_2005} also introduced ClassSheets, a system for visually developing object-oriented models of spreadsheets. In a series of related work, Cunha et al. \cite{giannakopoulou_type-safe_2011, cunha_embedding_2011, hutchison_bidirectional_2012} developed systems for bidirectionally syncing a ClassSheet model with a generated spreadsheet. While all of the spreadsheet modeling tools we have discussed aim to solve problems similar to those addressed by Kale, the tools are fundamentally different from the user's perspective. Model-driven development requires the user to interact with a separate representation of their spreadsheet, while Kale users interact with a single, unified representation. Model-driven development tools generate traditional spreadsheets with correctness properties--Kale aims to instead achieve these properties by introducing constraints on the structure of the spreadsheets and formulas with which users directly interact. Our work represents a first step into this design space; future work will be required to understand the usability difference between these paradigms.

\section{Limitations and Future Work}
Further study is needed to assess the real-world risks that structural changes pose in the context of spreadsheet \textit{authoring} tasks, beyond modification of existing spreadsheets. A more open-ended study could elucidate whether users tend to build spreadsheets that are fragile in the way ours were. The tasks in this study were designed to produce the associated risks; future work is required to measure how often these situations appear during real-world tasks. Additionally, while our corpus study demonstrates that replicating real spreadsheets is \textit{possible} in Kale, we would also like to evaluate how Kale's structural constraints impact real spreadsheet authors and viewers. The effect on users of splitting spreadsheets into separate, explicit tables is unclear, and warrants further investigation.

Kale is a prototype system. One key feature we did not evaluate in our study (due to an immature implementation) is \emph{queries}, which would enable formulas to reference a subset of a range. For example, \texttt{AVERAGE(Salary[Experience > 10])} might compute the average salary of employees with over 10 years of experience. Another feature that we did not evaluate is cross-table references, which in practice may be important for representing spreadsheets with more diverse structures.

\section{Conclusion}

In this paper, we showed evidence that structural changes in spreadsheets may present a common and serious risk to spreadsheet users. Kale represents an approach to eliminate these risks posed by traditional spreadsheet reference representation and semantics. Kale restricts the forms references can take and re-defines the \emph{absolute} and \emph{relative} distinction, allowing users to specify when writing formulas how they should be updated for structural changes. Evidence from our study suggests that adopting safer spreadsheet reference behavior may significantly reduce the incidence of spreadsheet errors.

\section{Declarations}

\subsection{Funding}
Not applicable.

\subsection{Ethical Approval}
All studies were approved by our institution's Institutional Review Board (IRB).

\subsection{Informed Consent}
All data collected from study participants was obtained with informed consent.

\subsection{Author Contributions}
Michael Coblenz and Jacob Yim contributed to system design and implementation, user study design, and paper writing.
Ajinkya Bokade, Mounika Padala, Julia Epshtein, Priyanka Bhatia, Piyush Chauhan, Simran Gill, and Grishma Gurbani contributed to system design and implementation.
Aniket Gupta, Vaibhav Khetan, Arushi Munjal, and Jeffery Tung contributed to the corpus study.
Joanna Yang, Mounika Padala, and Priyanka Bhatia contributed to the user study design.

\subsection{Data Availability Statement}
A replication package containing all data supporting the findings of our user study, as well as tasks and materials provided to participants, is available at the following URL: \url{https://github.com/ucsd-salad/kale-replication}. The Kale system is available as an open-source project at the following URL: \url{https://github.com/ucsd-salad/Kale}.

\subsection{Conflict of Interest}
Not applicable.





\bibliography{sn-bibliography}

\end{document}